\documentstyle[12pt]{article}

\begin{document}

\bibliographystyle{unsrt}

%---------------------------------------------------------------
%
\begin{center}
{\Large \bf\boldmath  Transition strengths from 
$^{\bf 10}$B$(e,e')^{\bf 10}$B}\\
\vspace{0.4cm}
{D. J. Millener}\\
\bigskip
 Brookhaven National Laboratory, Upton, NY 11973 \\
\end{center}
\begin{abstract}
 Inelastic electron scattering form factors are fitted with
polynomial times Gaussian expressions in the variable $y = (bq/2)^2$
to extract electromagnetic transition strengths at the photon point.
\end{abstract}
%\singlespaced
%---------------------------------------------------------------

\section{Introduction}

  The table of radiative widths from the 1979 Ajzenberg-Selove
tabulation was based on the low-$q$ results of Spamer \cite{spamer66}
(Darmstadt) for the 6.03-MeV $4^+;0$ level and the 7.48-MeV $2^+;1$ level,
togther with the $180^\circ$ results of Fagg et al. \cite{fagg76} (NRL)
for a number of levels. The 1984 and 1988 tabulations added results based 
on the work of Ansaldo et al. \cite{ansaldo79} (Saskatoon) for 
$0.61 < q < 1.81$ fm$^{-1}$ but did not take into account the erratum 
to that work \cite{ansaldo79}.

 The more recent work of Cichocki et al.~\cite{cichocki95} (NIKHEF)
gives longitudinal and transverse form factors in the range
$0.48 < q < 2.58$ fm$^{-1}$ for most levels up to the 6.56-MeV $4^-;0$
level. The analysis in this work includes extensive shell-model
calculations and the extraction of B(C2) values for five levels
of $^{10}$B. The analysis also includes data up to $q\sim 4$ fm$^{-1}$
taken at $180^\circ$ for the ground-state, the 1.74-MeV level, and the
5.17-MeV level \cite{hicks88} (Bates). For most transitions, the
form factors are plotted as a function of the effective momentum
transfer $q_{\rm eff} = q(1+2.75/E_0)$, where the beam energy $E_0$
is in MeV. This way of relating form factors in the plane-wave and 
distorted-wave Born approximations must also be applied to the data 
from the earlier works.  
 
 Cichocki et al. used a polymomial times Gaussian ($e^{-y}$) in the
variable $y = (bq/2)^2$, where $b$ is the harmonic oscillator
parameter, to represent the form factors and extract B(C2) values.
The proceedure is spelled out by Millener et al.~\cite{millener89}
who defined
\begin{eqnarray}
 B(C\lambda,q) = f^{-2}\, \frac{Z^2}{4\pi}\ 
\left[\frac{(2\lambda +1)!!}{q^\lambda} \right]\ F_L^2\ , \label{eq:C}\\
 B(M\lambda,q) = f^{-2}\, \frac{Z^2}{4\pi}\ \frac{\lambda}{\lambda +1}\ 
\left[\frac{(2\lambda +1)!!}{q^\lambda} \right]\ F_T^2 \ , \label{eq:M}
\end{eqnarray}
where $f= f_{\rm SN}f_{\rm c.m.}e^{-y}$ takes out the exponential
dependences in the (theoretical) form factor. In the conventional
definition of $B(C\lambda,q)$ and $B(M\lambda,q)$, we should set
$f=1$. Because nature does not know about $f_{\rm c.m.}$ (and even
in theory we don't need it if we use an appropiate system of 
relative coordinates), we perform the fit, with $f=f_{\rm SN}e^{-y}$, to
\begin{eqnarray}
 B(\lambda,q)^{1/2}& = & f\, (A + By + Cy^2 + ...)\label{eq:Ba} \\
  & = & \left[\frac{b}{2}\right]^\lambda\ (A' + B'y + C'y^2 + ...)\ .
\label{eq:Bb}
\end{eqnarray}
  
\section{Corrections for electron distortion}

 We use the effective momentum transfer $q_{\rm eff} = q(1+2.75/E_0)$
prescription \cite{cichocki95} to approximately correct the original 
data for electron distortion so that we can use form factors calculated 
in the plane-wave Born approximation. This is especially important
at low incident energies $E_0$ and needs to be performed for the
Darmstadt~\cite{spamer66}, NRL~\cite{fagg76}, and Saskatoon~\cite{ansaldo79}
data as sketched in the next subsections.

\subsection{Darmstadt data}
 
 The measured quantity is the ratio of inelastic to elastic cross
section and we could use this data together with a modern
parametrization of the elastic cross section. The derived quantity
$B(\lambda,q)$ is tabulated as a function of $q^2$ where
\begin{equation}
 q^2 = 2k_0^2\, (1- cos\theta)\, (1-k/k_0) + k^2    \label{eq:q}
\end{equation}
and $k=E_x/\hbar c$ and $k_0=E_0/\hbar c$ ($\hbar c = 197.32696$ MeV.fm).
We  recalculate $q$ and calculate $q_{\rm eff}$ using the
tabulated values of $E_0$ and $\theta$. The units for $B(\lambda,q)$
are given as $10^{-51}$ cm$^4 = 10$ fm$^4$ for C2 and  
$10^{-28}$ cm$^2 = 10^{-2}$ fm$^2$ for M1. We absorb the factor
of $\alpha\hbar c = e^2$ from the expressions of Spamer~\cite{spamer66}
so that $B(\lambda,q)$ is expressed in the conventional units of
$e^2$.fm$^{2\lambda}$. Then, from Eqs.~(\ref{eq:C}) and (\ref{eq:M})
(with $f=1$) in terms of the $B(\lambda,q)$ tabulated in 
Ref.~\cite{spamer66}
\begin{equation}
 F_L^2 = 10\, B(C2)\times 2.234\times 10^{-3}\times q^4 \label{eq:fl}
\end{equation}
for the longitudinal form factor of the 6.025-MeV $4^+;0$ level, and
\begin{equation}
 F_T^2 = 10^{-2}\, B(M1)\times 1.117\times 10^{-1}\times q^2 \label{eq:ft}
\end{equation}
for the transverse form factor of the 7.477-MeV $2^+;1$ level.

\subsection{NRL data}

 Fagg et al.~\cite{fagg76} give the $180^\circ$ cross sections in
nb/sr ($= 10^{-7}$ fm$^2$/sr) for three incident energies 
(40.5, 50.6, and 60.6 MeV and we have ($e^2 = 1.44$ MeV.fm)
\begin{equation}
\frac{d\sigma}{d\Omega} = \frac{Z^2\,e^4}{4\,E_0^2\,R}\ .\ F_T^2 
\label{eq:fagg}
\end{equation}
with the recoil factor $R=(1+2E_0/M)$ (M = nuclear mass, and  e.g.,
$R=1.013$ for $E_0 = 60.6$ MeV) and $q$ from Eq.~(\ref{eq:q}).
  
\subsection{Saskatoon data}

 The data are already tabulated as form factors and we simply change
$q$ to $q_{\rm eff}$.

\section{C2 transitions}

 In their appendix, Cichocki et al.~\cite{cichocki95} extract B(C2)
values for five states using Eq.~(\ref{eq:Ba}), perhaps without the 
inclusion of the single-nucleon form factor $f_{SN}$ but this is 
essentially irrelevant at the photon point. For the 6.025-MeV level,
the low-$q$ data from Darmstadt and the Saskatoon data were also
included. In the following subsections, we discuss fits to each level
starting with the 6.025-MeV $4^+;0$ level. As in Ref.~\cite{cichocki95},
the oscillator parameter is fixed at 1.60 fm. In principle, we could
include $b$ in the fit but it turns out that $b=1.60$ fm is close to the
optimum value. Besides a change of $b$ in Eq.~(\ref{eq:Bb}) is compensated
for by a change in $A'$ when fitting to  data. Of course,
theoretical B(C2) values calculated with harmonic oscillator wave functions
scale as $b^4$.

\subsection{The 6.025-MeV $4^+;0$ level}

 The original Darmstadt value for the B(C2) is $24.4\pm 2.5$ $e^2$fm$^4$
(the inclusion of 5 data points from Orsay lead to a slightly smaller
value of $23.4\pm 2.5$ $e^2$fm$^4$) - this value is quite well reproduced 
in the first line of Table~\ref{tab:602}. Taking the effect of distortion 
\begin{table}[b]
\caption{Fits to C2 form factors for the 6.025-MeV $4^+;0$ level using 
Eq.~(\ref{eq:Bb}). The harmonic oscillator parameter is fixed at $b=1.60$ fm.
The quantities in parentheses are standard deviations. When $\chi^2$/DF
is greater than one, the error on B(C2) is inflated by the square root of
this quantity.}
\begin{tabular*}{\textwidth}{@{}c@{\extracolsep{\fill}}ccccccc}
\hline
 Data & N & $\chi^2$/DF & $A$ & $B$ & $C$ & $D$ & B(C2) \\
\hline
 D $^a$ & 9 & 0.392 & 7.707(410) & $-2.18(231)$ & &  & 24.33(259) \\
 D $^b$ & 9 & 0.346 & 6.507(369) & \phantom{-}2.14(178) & &  & 17.34(197) \\
 D      & 9 & 0.365 & 6.746(139) & & &  & 18.64(77) \\
N+S     & 28 & 1.59 & 6.707(131) & $-0.945(290)$ & 0.435(173) & $-0.010(30)$ &
18.42(91)  \\ 
N+S  $^c$   & 25 & 1.69 & 6.600(194) & $-0.545(556)$ & 0.032(452) & 0.013(109)
 & 17.84(137)  \\ 
N+S $^c$    & 25 & 1.62 & 6.619(108) & $-0.607(194)$ & 0.085(71) &  &
17.95(74)  \\ 
N+S+D     & 37 & 1.28 & 6.761(106) & $-1.052(244)$ & 0.491(71) & -0.108(27) &
18.72(66)  \\ 
N+S+D $^c$    & 34 & 1.33 & 6.727(137) & $-0.875(417)$ & 0.275(360) &
 -0.040(90) & 18.53(87)  \\ 
N+S+D $^c$  & 34 & 1.30 & 6.682(91) & $-0.707(169)$ & 0.118(64) &  &
18.29(57) \\ 
\hline
\end{tabular*}

\vspace{0.5\baselineskip}
{\noindent \footnotesize
${}^{\rm a}$ Uncorrected Darmstadt data.\\
${}^{\rm b}$ Darmstadt data vs. $q_{\rm eff}$ with $q$ from $E_0$, $\theta$. \\
${}^{\rm c}$ $q_{\rm eff} < 2$ fm$^{-1}$.}
\label{tab:602}
\end{table}
into account via the $q_{\rm eff}$ prescription results in a considerably lower
value of $17.34\pm 1.97$ $e^2$fm$^4$ (note that B(C2) = $(b/2)^4\, A^2$)
as pointed out by Cichocki et al.~\cite{cichocki95} - taking the $q$
values from Table 1 of Spamer~\cite{spamer66} instead of recomputing
them gives B(C2) = $17.66\pm 1.98$ $e^2$fm$^4$. Note that the parameter
$B$ is not well determined and that the $e^{-y}$ term in the oscillator
form factor pretty much takes into account the terms involving the
transition radius in the original Darmstadt paper. As the third line of
Table~\ref{tab:602} shows one can obtain a one-parameter fit of similar
quality but with a smaller error because of the restrictive nature of
the fitting function. The fact that essentially a p-shell form factor
fits so well is surprising because the transition is very strong
and the higher-order terms responsible for this should lead to
a $B$ coefficient which is negative (e.g., the hamonic oscillator form
factor for the 2$\hbar\omega$ giant quadrupole resonance is of the form
$y(1-1/3y)e^{-y}$ and coherence at low $q$ means a negative coefficient
for the next term).

 If we fit the NIKHEF data using a 3-parameter polynomial, the $\chi^2$/DF
is 1.77; for the NIKHEF + Saskatoon data, it is 2.02. Adding an extra term
to take care of the high-$q$ behavior leads to some improvement (line
4 of Table~\ref{tab:602}). Removing the three data points with 
$q_{\rm eff} > 2$ fm$^{-1}$ doesn't lead to much change, although a
three-parameter fit is now possible, as the next two lines of 
Table~\ref{tab:602} show. The final three lines of Table~\ref{tab:602} show
fits to the complete data set. The four-parameter fit gives
B(C2) = $18.7\pm 0.7$ $e^2$fm$^4$. To compare with the electromagnetic
value for the $4^+\to gs$ transition, we multiply by 7/9 and convert
to Weisskopf units (1 W.u. = 1.2797 $e^2$fm$^4$) getting $11.4\pm 0.4$ W.u.
This agrees with the electromagnetic value of  $12.4\pm 1.8$  W.u.,
which is derived from the $\omega\gamma$ value from the 
$^6$Li$(\alpha,\gamma)$ reaction and an E2/M1 mixing ratio.

\subsection{The 0.718-MeV $1^+;0$ level}

 The lifetime for this long-lived level is precisely known, $\tau = 
1.020\pm 0.005$ nsec. This corresponds to a B(C2) for electron
scattering of 1.796(9) $e^2$fm$^4$. The value of 1.71(14)  $e^2$fm$^4$
in the first line of Table~\ref{tab:072} derived from the NIKHEF data 
is in good agreement. Therefore including the electromagnetic value
as a data point changes the $\chi^2$ only slightly. A three-parameter fit
gives a significant increase in $\chi^2$.

\begin{table}[hb]
\caption{Fits to C2 form factors for the 0.718-MeV $1^+;0$ level using 
Eq.~(\ref{eq:Bb}). The harmonic oscillator parameter is fixed at $b=1.60$ fm.
The quantities in parentheses are standard deviations. The error on
B(C2) is inflated by $\sqrt{\chi^2/DF}$.}
\begin{tabular*}{\textwidth}{@{}c@{\extracolsep{\fill}}ccccccc}
\hline
 Data & N & $\chi^2$/DF & $A$ & $B$ & $C$ & $D$ & B(C2) \\
\hline
N     & 14 & 1.29 & 2.045(73) & $-1.042(155)$ & 0.390(94) & $-0.058(17)$ &
1.71(14)  \\ 
N+EM     & 15 & 1.21 & 2.093(5) & $-1.144(51)$ & 0.450(47) & $-0.068(11)$ &
1.795(9) \\ 
\hline
\end{tabular*}

\label{tab:072}
\end{table}

\subsection{The 2.154-MeV $1^+;0$ level}

  The first line of Table~\ref{tab:215} shows a 3-parameter fit to all
the NIKHEF data points while the next line shows the same fit with the
highest $q$ data point removed. The 2-parameter fit in the third line shows 
very little deterioration in $\chi^2$. The last line shows a 1-parameter
fit which is still acceptable in terms of $\chi^2$ but is certainly not as 
good as the other fits. The $\chi^2$ doesn't change for $b=1.56$ fm or
$b=1.66$ fm and neither does B(C2) to any significant extent.

 The electromagnetic data in the current tabulation gives 0.75(9) 
$e^2$fm$^4$ for the B(C2) up. This depends on a number of values for
lifetime ($2.13\pm 0.20$ ps) and the ground-state branch ($21.1\pm 1.6$ \%).
Probably, the previous lifetime average of $2.30\pm 0.26$ ps should
be used but this only gets the the B(C2) down to 0.69 $e^2$fm$^4$ (the
lowest $\gamma$-ray branch of 17.5\% would give 0.57 $e^2$fm$^4$).

\begin{table}[ht]
\caption{Fits to C2 form factors for the 2.154-MeV $1^+;0$ level using 
Eq.~(\ref{eq:Bb}). The harmonic oscillator parameter is fixed at $b=1.60$ fm.
The quantities in parentheses are standard deviations.}
\begin{tabular*}{\textwidth}{@{}c@{\extracolsep{\fill}}ccccccc}
\hline
 Data & N & $\chi^2$/DF & $A$ & $B$ & $C$ & $D$ & B(C2) \\
\hline
N     & 13 & 0.84 & 0.963(50) & 0.091(74) & $-0.015(22)$ & & 0.380(36) \\ 
N $^a$ & 12 & 0.40 & 1.005(52) & 0.010(81) & 0.013(25) & & 0.413(43) \\ 
N $^a$ & 12 & 0.39 & 0.981(27) & 0.052(17) &           & & 0.394(22) \\ 
N $^a$ & 12 & 1.15 & 1.047(14) &           &           & & 0.449(14) \\ 
\hline
\end{tabular*}
\vspace{0.5\baselineskip}
{\noindent \footnotesize
${}^{\rm a}$ Highest $q$ data point removed.}
\label{tab:215}
\vspace*{-0.5cm}
\end{table}

\subsection{The 3.587-MeV $2^+;0$ level}

 The first line of Table~\ref{tab:359} shows a 3-parameter fit to all
the NIKHEF data points which yields B(C2) = $0.616\pm 0.044$ $e^2$fm$^4$
which is in reasonable agreement with the electromagnetic value of
 $0.85\pm 0.25$ $e^2$fm$^4$. The latter depends on lifetime, branch,
and mixing ratio.

\begin{table}[h]
\caption{Fits to C2 form factors for the 3.587-MeV $2^+;0$ level using 
Eq.~(\ref{eq:Bb}). The harmonic oscillator parameter is fixed at $b=1.60$ fm.
The quantities in parentheses are standard deviations.}
\begin{tabular*}{\textwidth}{@{}c@{\extracolsep{\fill}}ccccccc}
\hline
 Data & N & $\chi^2$/DF & $A$ & $B$ & $C$ & $D$ & B(C2) \\
\hline
N     & 16 & 1.16 & 1.226(42) & $-0.130(64)$ & 0.061(21) & & 0.616(44) \\ 
N     & 16 & 1.21 & 1.261(61) & $-0.226(139)$ & 0.129(91) & $-0.014(18)$
& 0.652(63) \\ 
\hline
\end{tabular*}
\label{tab:359}
\end{table}

\subsection{The 5.920-MeV $2^+;0$ level}

   The first line of Table~\ref{tab:592} shows a 3-parameter fit to all
the NIKHEF data points while the next line shows the same fit with the
highest $q$ data point removed. The 2-parameter fit in the third line shows 
very little deterioration in $\chi^2$. The same can be said of the 1-parameter
fit in the last line but the B(C2) changes from 0.164 to 0.202 as 
$b$ changes from 1.55 fm to 1.65 fm. In  2-parameter or 3-parameter fits
the $\chi^2$ and B(C2) vary little with modest changes in $b$. 
\begin{table}[hb]
\caption{Fits to C2 form factors for the 5.920-MeV $1^+;0$ level using 
Eq.~(\ref{eq:Bb}). The harmonic oscillator parameter is fixed at $b=1.60$ fm.
The quantities in parentheses are standard deviations.}
\begin{tabular*}{\textwidth}{@{}c@{\extracolsep{\fill}}ccccccc}
\hline
 Data & N & $\chi^2$/DF & $A$ & $B$ & $C$ & $D$ & B(C2) \\
\hline
N     & 12 & 0.89 & 0.555(91) & 0.279(182) & -0.157(83) & & 0.126(35) \\ 
N $^a$ & 11 & 0.87 & 0.602(100) & 0.161(213) & $-0.089(105)$ & & 0.148(49) \\ 
N $^a$ & 11 & 0.85 & 0.679(38) & $-0.015(44)$ &           & & 0.189(21) \\ 
N $^a$ & 11 & 0.78 & 0.667(13) &           &           & & 0.182(7) \\ 
\hline
\end{tabular*}
\vspace{0.5\baselineskip}
{\noindent \footnotesize
${}^{\rm a}$ Highest $q$ data point removed.}
\label{tab:592}
\end{table}

\section{M3 transitions}

 In addition to the transition to the 1.74-MeV $0^+;1$ level, the
transverse form factor to the 5.164-MeV $2^+;1$ level is dominantly
M3 with a small correction for M1 at low $q$. The $q_{\rm eff}$
prescription can be used on the Saskatoon data but the NIKHEF
data for the  $0^+;1$ level is given as a function of $q$ and
can't be corrected without a knowledge of $E_0$ for each point.
However, a B(M3) is available from a DWBA analysis of the data.
Note that for $A=10$, 1 W.u. = 35.548 $\mu^2$fm$^4$ = 0.3932 $e^2$fm$^6$.

\subsection{The 1.740-MeV $0^+;1$ level}

 We first note that $\Gamma_\gamma = (1.05\pm 0.25)\times 10^{-9}$ from
the original analysis of the Saskatoon data (see erratum of 
Ref.~\cite{ansaldo79}) corresponds to B(M3$\uparrow) = (8.27\pm 1.97)$
$e^2$fm$^6$ = $(748\pm 178)$ $\mu^2$fm$^4$.

 The first line of Table~\ref{tab:174} gives B(M3$\uparrow) = 
(804\pm 110)$ $\mu^2$fm$^4$ for a fit to the data as a function of
$q$. This is reduced to $(688\pm 101)$ $\mu^2$fm$^4$ for a fit to the 
data as a function of $q_{\rm eff}$. The value from a DWBA fit to
the complete data set shown in Ref.~\cite{cichocki95} is
633 $\mu^2$fm$^4$ (R. Hicks, private communication).

\begin{table}[h]
\caption{Fits to M3 form factor for the 1.740-MeV $0^+;1$ level using 
Eq.~(\ref{eq:Bb}). The harmonic oscillator parameter is fixed at $b=1.60$ fm.
The quantities in parentheses are standard deviations. The unit for B(M3)
is $e^2$fm$^6$. First line $q$, second line $q_{\rm eff}$.}
\begin{tabular*}{\textwidth}{@{}c@{\extracolsep{\fill}}ccccc}
\hline
 Data & N & $\chi^2$/DF & $A$ & $B$ &  B(M3$\uparrow$) \\
\hline
S     & 8 & 0.37 & 5.823(400) & 0.213(400) & 8.89(122) \\ 
S     & 8 & 0.36 & 5.386(396) & 0.522(388) & 7.61(112) \\ 
\hline
\end{tabular*}
\label{tab:174}
\end{table}

\subsection{The 5.164-MeV $2^+;1$ level}

The original analysis of the Saskatoon data \cite{ansaldo79} gave 
B(M3$\uparrow) = (21.6\pm 2.2)$ $e^2$fm$^6$ = $(1953\pm 19)$ $\mu^2$fm$^4$.
This fit included an M1 contribution.

 The first two lines of Table~\ref{tab:516} contain no correction for
the M1 contribution at low $q$. The first line fits the Saskatoon and
NIKHEF data as a function of $q_{\rm eff}$ while the second line also
contains the Catholic University of America low $q$ data. A significant
difference in the extracted B(M3) can be seen when the two points with
$q_{\rm eff} < 0.8$ fm$^{-1}$ are removed from the S+N data set (third line).

 Finally, we subtract an M1 contribution calculated by normalizing
the computed M1 shell-model form factor to the B(M1) obtained from
electromagnetic data. Because there is such a large cancellation
for the lowest $q$ data point of the CUA data set, we omit this point
entirely. This results in  B(M3$\uparrow) = 19.4\pm 2.0$ $e^2$fm$^6$ or
$(1756\pm 181)$ $\mu^2$fm$^4$; B(M3$\downarrow) = 27.2\pm 2.8$ $e^2$fm$^6$ or
$(69.1\pm 7.1)$ W.u. This corresponds to $\Gamma_\gamma = (1.00\pm 0.10)
\times 10^{-6}$ eV.

\begin{table}[h]
\caption{Fits to M3 form factor for the 5.164-MeV $2^+;1$ level using 
Eq.~(\ref{eq:Bb}). The harmonic oscillator parameter is fixed at $b=1.60$ fm.
The quantities in parentheses are standard deviations. The unit for B(M3)
is $e^2$fm$^6$.}
\begin{tabular*}{\textwidth}{@{}c@{\extracolsep{\fill}}cccccc}
\hline
 Data & N & $\chi^2$/DF & $A$ & $B$ & $C$ & B(M3$\uparrow$) \\
\hline
S+N     & 17 & 0.51 & 9.611(543) & $-3.059(945)$ & 1.608(364) & 24.2(27) \\ 
S+N+CUA  & 20 & 0.68 & 9.913(397) & $-3.571(726)$ & 1.793(294) & 25.8(21) \\ 
S+N $^a$ & 15 & 0.37 & 8.672(838) & $-1.569(1380)$ & 1.090(502) & 19.7(38) \\ 
S+N+CUA $^b$ & 19 & 0.39 & 8.612(452) & $-1.466(805)$ & 1.052(319) & 19.4(20) 
\\ 
\hline
\end{tabular*}
\vspace{0.5\baselineskip}
{\noindent \footnotesize
${}^{\rm a}$ $q_{\rm eff} > 0.81$ fm$^{-1}$. \\
${}^{\rm b}$ Theoretical $F^2_T$(M1) normalized to B(M1) = $0.023\pm 0.006$
W.u. subtracted and CUA $q=0.41$ fm$^{-1}$ point omitted because 
of a large cancellation.}
\label{tab:516}
\end{table}

\section{M1 transition for the 7.48-MeV level}

 Ansaldo et al. \cite{ansaldo79} give $\Gamma^0_\gamma = 11.75\pm 0.75$ eV
for this strong M1 transition while Spamer~\cite{spamer66} gives
$\Gamma^0_\gamma = 12.0\pm 2.2$ eV. However, Chertok~\cite{chertok69}
corrected the later value to $11.0\pm 2.2$ eV after distortion corrections
were taken into account. Note that 1 W.u. = 1.7905 $\mu^2$ =
0.0198 $e^2$fm$^2$.

 The fit as a function of $q_{\rm eff}$ in the first line of 
Table~\ref{tab:748} yields $\Gamma^0_\gamma = 10.84\pm 1.58$ eV.
Adding the CUA data points gives $\Gamma^0_\gamma = 11.00\pm 1.14$ eV.
The Saskatoon contains three points around the second maximum of the
M1 form factor. Adding these data points gives a worse fit and
$\Gamma^0_\gamma = 11.35\pm 0.37$ eV. Increasing the number of 
parameters to three improves the fit but gives a substantially larger
B(M1) value corresponding to $\Gamma^0_\gamma = 12.55\pm 0.58$ eV.

 As far as the B(M1) is concerned, it is preferable to stick with
the value derived from the low-$q$ data. 

\begin{table}[h]
\caption{Fits to M1 form factor for the 7.48-MeV $2^+;1$ level using 
Eq.~(\ref{eq:Ba}). The harmonic oscillator parameter is fixed at $b=1.60$ fm.
The quantities in parentheses are standard deviations. The unit for B(M1)
is $e^2$fm$^2$.}
\begin{tabular*}{\textwidth}{@{}c@{\extracolsep{\fill}}cccccc}
\hline
 Data & N & $\chi^2$/DF & $A$ & $B$ &  B(M1$\uparrow$) \\
\hline
D     & 12 & 0.56 & 0.133(10) & $-0.0126(945)$ &  & 0.0177(26) \\ 
D+CUA  & 15 & 0.54 & 0.134(7) & $-0.0123(45)$ &  & 0.0180(19) \\
D+CUA+S & 23 & 0.95 & 0.136(2) & $-0.162(5)$ &  & 0.0185(6) \\
D+CUA+S & 23 & 0.58 & 0.1432(33) & $-0.196(13)$ & 0.023(8) & 0.0205(10) \\
\hline
\end{tabular*}
\label{tab:748}
\end{table}

\section{C3 transitions}

 Cichocki et al.~\cite{cichocki95} present data on the form factors for the
$2^-$, $3^-$, and $4^-$ levels at 5.110 MeV, 6.127 MeV , and 6.561 MeV.
The longitudinal form factor for the isolated 6.56-MeV level is
best defined. The C3 Weisskopf unit is 5.94 $e^2$fm$^6$.

\subsection{The 6.56-MeV $4^-;0$ level}

 The C1 and C3 harmonic oscillator form factors for 1$\hbar\omega$ 
transitions cannot be distinguished. However, the shell-model
calculations in Ref.~\cite{cichocki95} indicate the the C3 transition is
dominant for the $4^-$ level.  

 The first line of Table~\ref{tab:656} shows a 4-parameter fit to the
full NIKHEF data set which shows that B,C, and D are not determined
and that there is a large error on B(C3). The second line shows a 
3-parameter fit and a significant change in B(C3) (but within errors).
Because we are interested in pinning down a low-$q$ parameter, the
third line shows the effect of omitting the two highest $q$ data 
points. Now C is undetermined and the final line shows a 2-parameter fit
to the reduced data set (there is no change in the overall $\chi^2$).
Then  B(C3$\uparrow$) = $21.8\pm 1.1$ $e^2$fm$^6$ and
B(C3$\downarrow$) = $17.0\pm 0.9$ $e^2$fm$^6$ = $2.9\pm 0.2$ W.u.

\begin{table}[h]
\caption{Fits to C3 form factor for the 6.560-MeV $4^-;0$ level using 
Eq.~(\ref{eq:Bb}). The harmonic oscillator parameter is fixed at $b=1.60$ fm.
The quantities in parentheses are standard deviations. The unit for B(C3)
is $e^2$fm$^6$.}
\begin{tabular*}{\textwidth}{@{}c@{\extracolsep{\fill}}ccccccc}
\hline
 Data & N & $\chi^2$/DF & $A$ & $B$ & $C$ & $D$ & B(C3$\uparrow$) \\
\hline
N     & 14 & 0.87 & 8.642(1180) & $-0.115(1880)$ & $-0.597(871)$ & 
 0.112(122) & 19.6(54) \\ 
N     & 14 & 0.86 & 9.621(479) & $-1.764(522)$ & $0.189(126)$ & 
  & 24.3(24) \\ 
N $^a$    & 12 & 0.89 & 9.115(651) & $-1.077(793)$ & $-0.004(210)$ & 
  & 21.7(31) \\ 
N $^a$    & 12 & 0.80 & 9.127(233) & $-1.093(121)$ &  &   & 21.8(11) \\ 
\hline
\end{tabular*}
\vspace{0.5\baselineskip}
{\noindent \footnotesize
${}^{\rm a}$ $q_{\rm eff} < 2.2$ fm$^{-1}$.}
\vspace*{-0.6cm}
\label{tab:656}
\end{table}

\subsection{The 6.13-MeV $3^-;0$ level}

 This form factor is not so well defined because of the difficulty
of separating the cross section from the strong $4^+$ level at
6.025 MeV. Again, the shell-model calculations of Ref.~\cite{cichocki95} 
indicate the the C3 transition is dominant but in this case a
significant C1 contribution is also predicted.

\begin{table}[b]
\caption{Fits to C3 form factor for the 6.130-MeV $3^-;0$ level using 
Eq.~(\ref{eq:Bb}). The harmonic oscillator parameter is fixed at $b=1.60$ fm.
The quantities in parentheses are standard deviations. The unit for B(C3)
is $e^2$fm$^6$.}
\begin{tabular*}{\textwidth}{@{}c@{\extracolsep{\fill}}cccccc}
\hline
 Data & N & $\chi^2$/DF & $A$ & $B$ & $C$ & B(C3$\uparrow$) \\
\hline
N     & 13 & 3.55 & 10.27(81) & $-1.30(134)$ & $-1.20(50)$ & 
 27.6(82) \\ 
N $^a$ & 11 & 2.85 & 11.21(38) & $-0.96(33)$ &  &  33.0(38) \\ 
N $^b$ & 9 & 1.54 & 11.25(39) & $-1.13(34)$ &  &  33.1(27) \\ 
\hline
\end{tabular*}
\vspace{0.5\baselineskip}
{\noindent \footnotesize
${}^{\rm a}$ $q_{\rm eff} < 1.7$ fm$^{-1}$. \\
${}^{\rm b}$ Points at $q_{\rm eff} = 1.08$ and 1.46 fm$^{-1}$ omitted.}
\label{tab:613}
\end{table}
 The first line of Table~\ref{tab:613} shows a 3-parameter fit to the
full NIKHEF data set which is poor and gives a large error on B(C3).
The second line shows the effect of omitting the two highest $q$ data 
points and reducing the number of parameters by one. A better $\chi^2$
is obtained in the last line by omitting two high data points.
Then  B(C3$\uparrow$) = B(C3$\downarrow$) = $33.1\pm 2.7$ $e^2$fm$^6$ 
= $5.6\pm 0.5$ W.u. 

\subsection{The 5.11-MeV $2^-;0$ level}

 Here, the shell-model calculations of Ref.~\cite{cichocki95} 
indicate that the C1 transition is dominant over C3. In addition,
there exists a non-zero B(E1$\downarrow$) of $(5.0\pm 1.0)
\times 10^{-4}$ W.u. that arises from isospin mixing. This
corresponds to B(E1$\uparrow$) = $(1.07\pm 0.21)\times 10^{-4}$ $e^2$fm$^2$. 

 The first line of Table~\ref{tab:511} assumes good isopsin and
therefore no A coefficient. Allowing A to be non-zero improves the
fit (second line). Including the photon point in the fit worsens the
$\chi^2$ somewhat but still gives a reasonable fit.

\begin{table}[h]
\caption{Fits to the longitudinal form factor for the 5.110-MeV $2^-;0$ 
level using Eq.~(\ref{eq:Ba}). The harmonic oscillator parameter is 
fixed at $b=1.60$ fm. The quantities in parentheses are standard 
deviations.}
\begin{tabular*}{\textwidth}{@{}c@{\extracolsep{\fill}}cccccc}
\hline
 Data & N & $\chi^2$/DF & $A$ & $B$ & $C$ & B(C1$\uparrow$) \\
\hline
N     & 8 & 1.62 &  & $0.259(11)$ & $-0.100(13)$ & 
0.0 \\ 
N     & 8 & 0.83 & $-0.036(16)$ & $0.366(47)$  & $-0.171(33)$ & 
 $(1.3\pm 1.1)\times 10^{-3}$ \\ 
N  $^a$ & 9 & 1.16 & $-0.0106(11)$ & $0.290(12)$  & $-0.121(13)$ & 
 $(1.09\pm 0.23)\times 10^{-4}$ \\ 
\hline
\end{tabular*}
\vspace{0.5\baselineskip}
{\noindent \footnotesize
${}^{\rm a}$ Including electromagnetic data for the photon point.}
\label{tab:511}
\end{table}

\section{Higher levels}

 Ansaldo et al.~\cite{ansaldo79} give a longitudinal form factor
for a level at 8.07 MeV with a width of 760 keV, which they assign as 
$2^+;0$, and a transverse form factor for the $2^+;1$/$3^-;1$ doublet at 
8.9 MeV. Fitting the data for the 8.07-MeV level yields a 
B(C2$\uparrow) = 5.1\pm 0.7$ $e^2$fm$^4$. This is about a quarter
of the strength of the very strong transition to the 6.025-MeV $4^+$
level. As Zeidman et al.~\cite{zeidman88} note, this should lead to
a very strong excitation in inelastic pion scattering which is not seen.
In fact, even adding the cross sections for states at 7.8 and 8.07 MeV
still gives less than 25\% of the expected cross section for an isoscalar
C2 excitation.
 
  For the 8.9-MeV doublet, it is difficult to say anything much without
guidance from the shell-model as to the dominant multipoles expected.
The state is stronger than expected for an isovector excitation in
inelastic pion scattering~\cite{zeidman88}.

This work was supported by the U.S.  Department of Energy under
Contract No. DE-AC02-98CH10886.

\end{document}